\documentclass[showpacs,aps,prb,reprint,twocolumn]{revtex4-1}

\usepackage{commath}
\usepackage{amsmath}
\usepackage{graphicx}
\usepackage{hyperref}
\usepackage{txfonts}

\draft 

\begin{document}


\title{Beating plasmonic losses with an intrinsic channel of gain: the cases with Ag and Al} 


\author{Hai-Yao Deng}
\email{h.deng@exeter.ac.uk}
\affiliation{School of Physics, University of Exeter, EX4 4QL Exeter, United Kingdom}

\begin{abstract} 
An elementary approach is employed to show that genuine surface effects could destabilize surface plasma waves (SPWs) supported on the interface between a metal and a dielectric (usually the vacuum) and give rise to an intrinsic channel of gain for these waves. A comprehensive SPW theory is presented taking into account both the inter-band transition effects and the dielectric effects. Experimental consequences, especially in regard to the possibility of overcompensating for the energy losses suffered by SPWs, are exemplified for two common plasmonic materials: silver (Ag) and aluminum (Al). In both metals, the intrinsic channel of gain is shown having substantially reduced but insufficient to overcompensate for the losses, for different reasons. In Ag, the inter-band transition effects significantly weaken the intrinsic gain channel and make it unable to overcompensate for the losses, while in Al it is because the loss rate is too big. Nevertheless, we find it possible to enhance the intrinsic gain rate by replacing the vacuum with a moderate dielectric so that the losses can be overcompensated in Ag. This prediction is ready for experimental exploitation. 
\end{abstract}


\maketitle 

\section{Introduction}
\label{sec:1}
\textit{Introduction}. Surface plasma waves (SPWs) -- electron density undulations existing on metal surfaces -- has been tooted as the most promising candidate enabler of nano photonics~\cite{barns2003}. One of the fundamental issues that have so far constrained SPWs from fledging rests with energy losses~\cite{khurgin}. Losses are ubiquitous in metal optics and occur by several channels, with Joule heat, Landau damping and inter-band absorption being the representative ones~\cite{khurgin}. It is widely believed that such losses are intrinsic and cannot be eliminated without the addition of external gain medium~\cite{khurgin,oulton,stockman2008,premaratne2017}. They would ultimately limit the functioning of SPWs in many applications~\cite{khurgin,oulton}. 

However, through recent works a drastically different picture has emerged. In a series of publications~\cite{deng2017a,deng2017b,deng2016}, we showed that SPWs possess a universally protected intrinsic channel of gain, which could be harnessed to counteract the losses and possibly compensate for them all at once. This channel of gain signifies a potential instability of the Fermi sea. It arises from surface-caused symmetry-breaking effects that are utterly beyond the conventional discourses of SPWs on the basis of the frequently used but inadequate hydrodynamic/Drude model~\cite{ritchie1957,harris1971,feibelman1982,fetter1986,raether1988,zayats2005,maier2007,pitarke2007,sarid2010,pendry2013,luo2014}. We used Boltzmann's theory to capture such effects, while noting two drawbacks. Firstly, this theory does not provide a clear physical picture of the electronic motions underlying the surface effects. Secondly, it does not account for the effects due to inter-band transitions, which are known to be very pronounced in -- for example -- noble metals~\cite{johnson1972}.  

The purpose of this work is two-fold. Firstly, we provide an elementary approach to the surface effects and thereby furnish a clear physical explanation of the existence of the intrinsic channel of gain. This approach is equivalent to Boltzmann's theory but based on simple solutions to the semi-classical equation of dynamics for electrons. Secondly, we clarify inter-band transition effects and the effects of replacing the vacuum with a dielectric. The theory is then applied to real materials: silver (Ag) and aluminum (Al). For Ag, we find that inter-band transitions significantly weaken the intrinsic channel of gain because of screening and inter-band absorption. SPWs on pristine Ag surface are then not helped by the channel. However, by replacing the vacuum with a dielectric, we find that the intrinsic channel can be considerably fortified at long wavelengths. As a result, plasmonic losses in Ag could possibly be completely compensated by simply tuning the system toward a critical temperature $T^*$. The situation with Al is qualitatively different. Here inter-band effects are weak but discernible. Although the channel of gain is robust, the loss rate for Al is much bigger and consequently, SPWs on pristine Al surface are invariably lossy. Topping a dielectric, in contrast to the case with Ag, only slightly strengthen the channel and turns out to be insufficient to make up for the losses. This prediction is ready for experimental exploration with routine techniques in plasmonics. 

\begin{figure*}
\begin{center}
\includegraphics[width=0.98\textwidth]{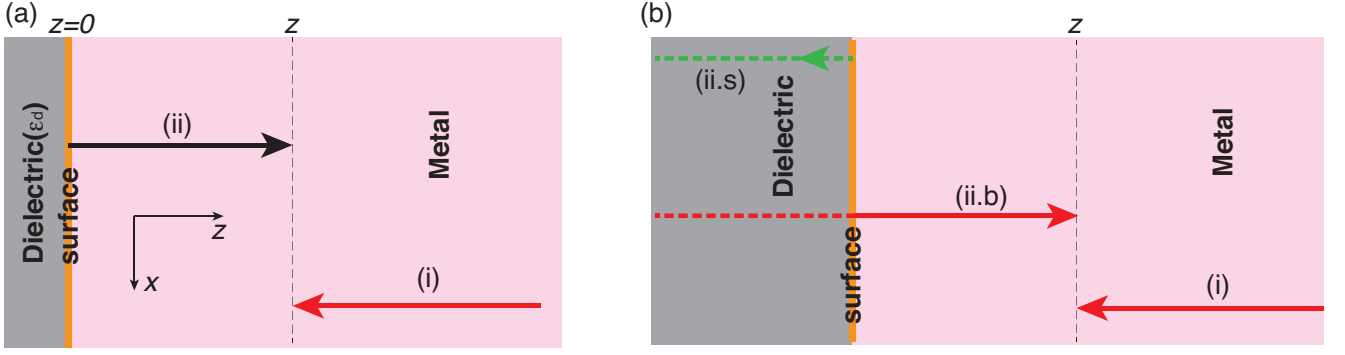}
\end{center}
\caption{ Illustration of the surface effects on the electrical responses of semi-infinite metals interfacing with a dielectric of dielectric constant $\epsilon_d$. Arrowed lines indicate the types of electronic paths (projected onto the z-axis). The surface is assumed to thermalize all incident electrons [following path (i) beginning at $z\rightarrow\infty$ with $v_z<0$] and no elastically reflected electrons are considered. Thermally emanated electrons, which follow path (ii) starting at the surface with $v_z>0$, from the surface can be decomposed into two virtual paths (ii.b) and (ii.s) [see Eqs.~(\ref{7}) and (\ref{8})]. Surface effects arise from (ii.s) while (i) and (ii.b) contain all bulk effects -- effects that are the same as in boundless systems. These surface effects lie beyond the scope of the hydrodynamic/Drude theory. They are responsible for the peculiar properties of surface plasma waves (SPWs) discussed in this work. Electronic motions by paths (ii.s) then give rise to the current density $\mathbf{J}_s(\mathbf{x},t)$ while the rest to $\mathbf{J}_b(\mathbf{x},t)$, see Eq.~(\ref{12}). Inter-band electronic transitions are not included here (see Sec.~\ref{sec:3}). \label{figure:f1}}
\end{figure*} 

\textit{Overview}. An overview of the paper is as follows. In the next section, we discuss intra-band electronic motions in the presence of SPWs supported on the surface of a semi-infinite metal (Fig.~\ref{figure:f1}). On the basis of semi-classical dynamics we evaluate the velocity corrections to individual electronic motions as a result of the SPWs. Such corrections are shown decomposable into a bulk component and a surface component. The former proves essentially the same as for a boundless bulk system whereas the latter represents genuine surface effects, which would totally vanish in a bulk system. A very general observation is that, the velocity corrections would appear singular unless an intrinsic channel for the SPWs exists. This observation is borne out by the actual solutions to the SPW equation established in Sec.~\ref{sec:4}. The velocity corrections can be used to obtain the electrical responses of the system, which of course can be accordingly split into a bulk component and a surface component. 

In  Sec.~\ref{sec:3}, the effects of inter-band transitions are discussed in terms of the concept of polarization current in the atomic limit. We prescribe a phenomenological expression for this current, which in combination with measurements or other input (e.g. \textit{ab initio}  calculations) can in principle be used to capture the inter-band effects. Along the line of our previous work~\cite{deng2017a,deng2017b}, we proceed to derive the key equation for SPWs supported on the interface between a metal and a general dielectric in Sec.~\ref{sec:4}. We discuss various limits of the equation and reveal some generic properties highlighting the inter-band and dielectric effects. We then solve the equation numerically and confirm the existence of the intrinsic channel of gain (Figs.~\ref{fig:f2} and \ref{fig:f3}). We show that the presence of a dielectric can enhance the channel. 

To demonstrate the possibility of beating the losses with the intrinsic channel alone in reality, we examine SPWs in two common plasmonic metals -- Ag and Al -- in Sec.~\ref{sec:5}. We use the fitting functions constructed by Raki\'{c} \textit{et al.}~\cite{rakic1998,rakic1995} to compute the polarization currents in these two materials and estimate the properties of SPWs in them (Fig.~\ref{fig:f4}). Some caveats of this approach are discussed therein. We find that in Ag the inter-band effects are pronounced and significantly weaken the intrinsic channel and as a result, SPWs on pristine Ag are always lossy. With a dielectric, however, that channel can be sufficiently fortified to beat all the losses that would plague the SPWs in this metal. An experimental scheme is proposed for this purpose (Fig.~\ref{fig:f5}). As for Al, in which inter-band effects are much less, we observe a strong intrinsic channel that yet still falls short of the huge losses present in this metal. A dielectric similarly enhances the channel but turns out to be insufficient to reverse the situation. We call for experiments to be carried out to investigate this prediction. 

We discuss the results and conclude the paper in Sec.~\ref{sec:6}, where evidences for the existence of the intrinsic channel of gain are also discussed. 

\section{Intra-band Electronic motions under SPWs}
\label{sec:2}
\subsection{Solutions of the semi-classical equation of motion}
We consider a semi-infinite metal lying in the region $z\geq0$ and bounded by a surface $z=0$. The other side of the space is the vacuum or more generally a dielectric with dielectric constant $\epsilon_d$, see Fig.~\ref{figure:f1}. We for the moment focus on intra-band electronic motions and the inter-band transitions will be discussed in the next section. As usual, the electrons of mean density $n_0$ are embedded in a background of uniformly distributed positive charges (the jellium model~\cite{pines}) keeping the system neutral. The conduction band where the electrons reside is assumed having a quadratic dispersion, i.e. the electrons' band energy $\varepsilon$ relates to its velocity $\mathbf{v}$ by $\varepsilon = m\mathbf{v}^2/2$, where $m$ denotes the effective electron mass. Retardation effects and magnetic fields are neglected. 

The semi-classical dynamic equations are invoked to describe the electronic motions under the electric field $\mathbf{E}(\mathbf{x},t)=\text{Re}\left[\mathbf{E}(z)e^{i(kx-\omega t)}\right]$ produced by the charge density $\rho(\mathbf{x},t) = \text{Re}\left[\rho(z)e^{i(kx-\omega t)}\right]$ due to the presence of SPWs. Here we have prescribed a quasi-plane wave form for the field quantities, with $k>0$ being the wavenumber, $\omega = \omega_s+i\gamma$ the complex frequency, $\mathbf{x} = (x,y,z)$ the spatial coordinates and $t$ the time. In addition, Re takes the real part of a quantity. It goes without saying that $\omega$ -- and hence the SPW frequency $\omega_s$ and its amplification/decay rate $\gamma$ -- is determined by the dynamics of the system (see Sec.~\ref{sec:4}). The equations read 
\begin{equation}
\frac{d}{dt}~\mathbf{X} = \mathbf{V}, \quad \left(\frac{d}{dt}+\frac{1}{\tau}\right)\mathbf{V} = (e/m)~\mathbf{E}\left(\mathbf{X}(t),t\right),\label{1}
\end{equation}
where $e$ is the electron charge and $\tau^{-1}$ gives the electronic collision rate. We have used $\mathbf{X}(t)$ and $\mathbf{V}(t)$ to denote the position and velocity at moment $t$ of the electrons. Without the electric field, an electron passing point $\mathbf{x}_0$ at instant $t_0$ would arrive at $\mathbf{x}$ at later instant $t$ via a rectilinear path with constant velocity 
\begin{equation}
\mathbf{v} = (\mathbf{x}-\mathbf{x}_0)/(t-t_0).
\label{2}
\end{equation}
The electric field distorts the path and corrects the velocity. To the first order in $\mathbf{E}(\mathbf{x},t)$, the velocity correction $\delta\mathbf{v} = \mathbf{V}-\mathbf{v}$ can be calculated without the path distortion. Under this approximation, $\delta\mathbf{v}$ can be obtained as
\begin{equation}
\delta\mathbf{v}(t) = \delta\mathbf{v}(t_0)e^{-\frac{t-t_0}{\tau}} + \frac{e}{m}\int^t_{t_0}dt' e^{-\frac{t-t'}{\tau}} \mathbf{E}\left(\mathbf{x}_0+(t'-t_0)\mathbf{v},t'\right). \label{3}
\end{equation}
With Eq.~(\ref{2}), the integral here over time can be transformed into one over spatial coordinates. Using the quasi-plane wave form for the electric field and $t-t' = (z-z')/v_z$ with $z' = z_0+(t'-t_0)v_z$ indicating a point on the rectilinear path, we find
\begin{equation}
\delta \mathbf{v}(t) = \delta\mathbf{v}(t_0)e^{-\frac{z-z_0}{v_z\tau}} + \text{Re}\left[e^{i(kx-\omega t)}\int^z_{z_0}\frac{dz'}{v_z} e^{i\frac{\tilde{\omega}(z-z')}{v_z}}\frac{e~\mathbf{E}(z')}{m}\right], \label{4}
\end{equation}
where $\tilde{\omega} = \bar{\omega} - kv_x$ with $\bar{\omega} = \omega + i/\tau$. This expression contains all needed to calculate the electrical responses of metals in the linear response regime. For later use, let us introduce $\bar{\omega} = \omega_s+i\gamma_0$ so that $\gamma = \gamma_0-1/\tau$. In the rest of this section, we show that $\gamma_0$ must always be non-negative, thereby warranting the existence of an intrinsic channel of gain for SPWs. 

Surface effects on electronic motions can in principle be accounted for by adding a surface field $\mathbf{E}_s$ to $\mathbf{E}$ in Eq.~(\ref{4}). However, the form of $\mathbf{E}_s$ is generally complicated and unknown \textit{a priori}. Further, $\mathbf{E}_s$ is not a single-valued function of materials: it can strongly depend on the fabrication process. Here we adopt a practical approach, observing that $\mathbf{E}_s$ acts only when electrons reach the surface and elsewhere it is irrelevant. Let us consider a plane located at $z$ parallel to the surface. The electrons arriving at this plane obviously fall in two groups depending on the sign of their $v_z$, see Fig.~\ref{figure:f1} (a). Electrons with $v_z<0$ move toward the surface but have not yet reached it. We may take in Eq.~(\ref{3}) $t_0$ to indicate the distant past, corresponding to $z_0\rightarrow\infty$. As SPWs are localized about the surface, we then expect for these electrons $\delta\mathbf{v}(t_0) = 0$, i.e. far away from the surface the electrons are not disturbed. According to Eq.~(\ref{4}), the velocity correction for them is then given by
\begin{equation}
\delta\mathbf{v}_{<}(\mathbf{x},\mathbf{v},t) = \text{Re}\left[e^{i(kx-\omega t)}\int^{z}_{\infty}\frac{dz'}{v_z} e^{i\frac{\tilde{\omega}(z-z')}{v_z}}\frac{e~\mathbf{E}(z')}{m}\right], ~ v_z<0.\label{5}
\end{equation}
Here we have underlined the explicit dependence of $\delta\mathbf{v}$ on $\mathbf{x}$ and $\mathbf{v}$ by writing them in as arguments. As they have not reached the surface yet, these electrons are obviously not affected by it at all. They move in the same manner as in a bulk system and therefore contribute to bulk electrical responses. 

The electrons with $v_z>0$ can have different stories. Needless to say, these electrons can be traced back to the electrons with $v_z<0$ which had already reached the surface and then been scattered by it. They are a direct consequence of the surface. In the simplest surface scattering picture, we may assume that incident electrons are either elastically reflected back or become thermalized. In this paper, we are not considering the elastically reflected electrons, although they can be easily handled and have already been studied in previous work~\cite{deng2016,deng2017a,deng2017b}. The thermalized electrons have no memory of their past and could start over with a rectilinear path beginning at the surface at any instant $t_0$. Thus, $\delta\mathbf{v}(t_0) = 0$ and $z_0 = 0$. The velocity correction then becomes
\begin{equation}
\delta\mathbf{v}_{>}(\mathbf{x},\mathbf{v},t) = \text{Re}\left[e^{i(kx-\omega t)}\int^z_{0}\frac{dz'}{v_z} e^{i\frac{\tilde{\omega}(z-z')}{v_z}}\frac{e~\mathbf{E}(z')}{m}\right], ~ v_z>0.\label{6}
\end{equation}
In both Eqs.~(\ref{5}) and (\ref{6}), we must have Im$(\tilde{\omega})\geq 0$. Otherwise, the integrals would diverge. This point will become more transparent in what follows. As such, we conclude that $\gamma_0$ must be non-negative, as asserted above and in our previous work and to be confirmed later on (Sec.~\ref{sec:4}).

In order to calculate the electrical current density from $\delta\mathbf{v}(\mathbf{x},\mathbf{v},t)$, we use the fact that the semi-classical electronic distribution function can be written as $f(\mathbf{x},\mathbf{v},t) = f_0\left(\varepsilon(\mathbf{v})\right) + g(\mathbf{x},\mathbf{v},t)$, where the equilibrium distribution $f_0$ is taken to be the Fermi-Dirac function, and in the regime of linear responses
\begin{equation}
g(\mathbf{x},\mathbf{v},t) = - \delta\mathbf{v}(\mathbf{x},\mathbf{v},t)\cdot\partial_{\mathbf{v}}f_0 = m\mathbf{v}\cdot\delta\mathbf{v}(\mathbf{x},\mathbf{v},t)(-f'_0) \label{11}
\end{equation}
with $f'_0 = \partial_{\varepsilon}f_0(\varepsilon)$. Expression (\ref{11}) is identical with  what was obtained before [Eq.~(A5) in Ref.~\cite{deng2017a}]. 

\subsection{Decomposition into bulk and surface components} 
Here we show that $\delta\mathbf{v}$ can be split into a bulk component $\delta\mathbf{v}_b$ and a surface component $\delta\mathbf{v}_s$, i.e. $\delta\mathbf{v} + \delta\mathbf{v}_s$, where $\delta\mathbf{v}_b$ has the same form as in a boundless system whereas $\delta\mathbf{v}_s$ exists only in the presence of a surface. We can then rewrite $g(\mathbf{x},\mathbf{v},t) = g_b(\mathbf{x},\mathbf{v},t) +  g_s(\mathbf{x},\mathbf{v},t) $, where $g_{b/s}$ follows from Eq.~(\ref{11}) with $\delta\mathbf{v}$ replaced by $\delta\mathbf{v}_{b/s}$. The corresponding electrical current density then also has a bulk and surface component, denoted by $\mathbf{J}_b$ and $\mathbf{J}_s$, respectively. They are given by
\begin{equation}
\mathbf{J}_{b/s}(\mathbf{x},t) = \left(\frac{m}{2\pi\hbar}\right)^3 \int d^3\mathbf{v}~e\mathbf{v}~g_{b/s}(\mathbf{x},\mathbf{v},t). \label{12}
\end{equation}
The expressions of $g_{b/s}(\mathbf{x},\mathbf{v},t)$ are complicated but can be found straightforwardly. They are not given in this paper. The expressions of $\mathbf{J}_{b/s}(\mathbf{x},t)$ are given in the next section. Needless to say, we can write $\mathbf{J}_{b/s}(\mathbf{x},t) = \text{Re}\left[e^{i(kx-\omega t)}\mathbf{J}_{b/s}(z)\right]$. This decomposition has also been established in previous work.~\cite{note1}

In previous work~\cite{deng2017a,deng2017b}, we carried out the decomposition in the following way. We first cast $\mathbf{E}(z)$ in terms of $\rho_q = \int^{\infty}_0dz\cos(qz)\rho(z)$ with the laws of electrostatics. We have $\mathbf{E}(z) = -\nabla \phi(z)$, where $\nabla = (ik,\partial_y,\partial_z)$ and the electrostatic potential is given by 
\begin{equation}
\phi(z) = (2\pi/k)\int^{\infty}_{-\infty} dz' ~e^{-k\abs{z-z'}}\left(\rho(z')+\rho_\text{d}(z')\right). \label{19}
\end{equation}
Here we have included the mirror charges $\rho_d(z)$ induced in the dielectric $\epsilon_d$ existing in the half space $z<0$. One can show that $$\rho_d(z\leq0) = -\frac{\epsilon_d-1}{\epsilon_d+1}~\rho(-z).$$ Performing the integration in Eq.~(\ref{19}), we obtain for $z\geq0$ 
\begin{equation}
E_x(z) = -i\int^{\infty}_0 dq~\frac{4k~\rho_q}{k^2+q^2}\left(2\cos(qz)-\beta e^{-kz}\right), \label{a1}
\end{equation}
where $\beta = 2\epsilon_d/(\epsilon_d+1)$, and 
\begin{equation}
E_z(z) = \int^{\infty}_0 dq~\frac{4k~\rho_q}{k^2+q^2}\left(2(q/k)\sin(qz)-\beta e^{-kz}\right). \label{a2}
\end{equation}
Next we substitute these expressions in Eqs.~(\ref{5}) and (\ref{6}) and do the integration over $z'$. We will end up expressions for $\delta\mathbf{v}_{>/<}$ that contain terms headed by functions $\cos(qz), \sin(qz), e^{-kz}$  and $e^{i\frac{\tilde{\omega}z}{v_z}}$. We collect those terms with $e^{i\frac{\tilde{\omega}z}{v_z}}$ in $\delta\mathbf{v}_s$ while the remaining in $\delta\mathbf{v}_b$, thus achieving the decomposition. 

However, we can also achieve the same decomposition using a more direct approach. To this end, we break down expression~(\ref{6}) according to the identity that $\int^z_0 = \int^{-\infty}_{0}+\int^{z}_{-\infty}$, as illustrated in Fig.~\ref{figure:f1} (b), so that $\delta\mathbf{v}_{>}(\mathbf{x},\mathbf{v},t) = \delta\mathbf{v}^{(1)}(\mathbf{x},\mathbf{v},t)+\delta\mathbf{v}^{(2)}(\mathbf{x},\mathbf{v},t)$, with
\begin{equation}
\delta\mathbf{v}^{(1)}(\mathbf{x},\mathbf{v},t) = \text{Re}\left[e^{i(kx-\omega t)}\int^z_{-\infty}\frac{dz'}{v_z} e^{i\frac{\tilde{\omega}(z-z')}{v_z}}\frac{e~\mathbf{E}(z')}{m}\right], \label{7}
\end{equation}
which would represent exactly the counterpart of $\delta\mathbf{v}_<(\mathbf{x},\mathbf{v},t)$ for electrons moving toward the point $\mathbf{x}$ from $z\rightarrow-\infty$ in boundless metals, and
\begin{equation}
\delta\mathbf{v}^{(2)}(\mathbf{x},\mathbf{v},t) = \text{Re}\left[e^{i(kx-\omega t)}\int^{-\infty}_{0}\frac{dz'}{v_z} e^{i\frac{\tilde{\omega}(z-z')}{v_z}}\frac{e~\mathbf{E}(z')}{m}\right].
\label{8}
\end{equation} 
In this mathematical transformation, the electric field on the dielectric side $z<0$ is 'fictitious' and does not have any impact on the electronic motions on the metal side. Thus, the form of $\mathbf{E}(z<0)$ is physically irrelevant and can in principle take on any values at this stage. We can now combine $\delta\mathbf{v}_<$ and $\delta\mathbf{v}^{(1)}$ in a single form $\delta\mathbf{v}_b(\mathbf{x},\mathbf{v},t) = \text{Re}\left[ e^{i(kx-\omega t)} \delta\mathbf{v}_b(z,\mathbf{v})\right]$, where 
\begin{eqnarray}
\delta\mathbf{v}_b(z,\mathbf{v}) &=& \int^{\infty}_{-\infty}\frac{dz'}{\abs{v_z}} e^{i\tilde{\omega}\abs{\frac{z-z'}{v_z}}}\frac{e~\mathbf{E}(z')}{m} \nonumber \\ &~&~~~~~~\times \left(\Theta(v_z)\Theta(z-z')+\Theta(-v_z)\Theta(z'-z)\right). \label{9}
\end{eqnarray}
The remaining $\delta\mathbf{v}^{(2)}$ arises only when the surface is present and is then identified with $\delta\mathbf{v}_s$. Writing $\delta\mathbf{v}_s(\mathbf{x},\mathbf{v},t) = \text{Re}\left[ e^{i(kx-\omega t)} \delta\mathbf{v}_s(z,\mathbf{v})\right]$, we find
\begin{equation}
\delta\mathbf{v}_s(z,\mathbf{v}) = - ~\Theta(v_z)~\int^{0}_{-\infty}\frac{dz'}{v_z} e^{i\frac{\tilde{\omega}(z-z')}{v_z}}\frac{e~\mathbf{E}(z')}{m}\propto e^{i\frac{\tilde{\omega}z}{v_z}}, \label{10}
\end{equation}
To ensure that Eqs.~(\ref{9}) and (\ref{10}) indeed produce the same results as obtained using the method as recapitulated above of Ref.~\cite{deng2017a,deng2017b}, we impose that the fictitious electric field on the side $z<0$ has the same form as $\mathbf{E}(z\geq0)$ but with $k$ adiabatically vanishing in the limit $z\rightarrow-\infty$. 

Equations (\ref{9})  and (\ref{10}) obviously demand that $\gamma_0=\text{Im}(\tilde{\omega})\geq 0$, in support of what is said above.

\section{Inter-band transition effects}
\label{sec:3}
The current density as given by Eq.~(\ref{12}) only accounts for electronic motions in the conduction energy band. In general, electrons residing in lower-energy bands can also contribute by means of virtual transitions to the conduction band. As they are tightly held to host atoms, these electrons are usually referred to as 'bound electrons'. The current density due to them is no more than the polarization current $\mathbf{J}_p$. One may employ Kubo's formula to compute $\mathbf{J}_p$, which is however material specific and often a formidable task. Here we take a phenomenological approach, observing that lower-energy bands are largely dispersionless and hence barely susceptible to spatial terminations. The metal may be approximated as a collection of loosely bonded atoms -- as in an insulator -- in regard to the bound electrons, so that $\mathbf{J}_p$ can be obtained as a sum of the contributions from individual atoms. As such, we may write 
\begin{equation}
\mathbf{J}_p(\mathbf{x},t) = \text{Re}\left[e^{i(kx-\omega t)}\sigma_{p}(\omega)\mathbf{E}(z)\right], \label{13}
\end{equation}
with the inter-band conductivity $\sigma_p(\omega)$ taking the bulk values. 

Note that $\sigma_p(\omega)$ can be computed in the atomic limit and usually modeled in the Lorentz form. It is related to the inter-band dielectric response by $\epsilon_p(\omega) = 4\pi i\sigma_p(\omega)/\omega$, which can be measured using for example ellipsometry. $\epsilon_p(\omega)$ contains a real part $\epsilon_{pr}(\omega)$ and an imaginary part $\epsilon_{pi}(\omega)$. While $\epsilon_{pr}(\omega)$ acts to shield the conduction electrons, $\epsilon_{pi}(\omega)$ -- which is always positive -- leads to inter-band absorption. These effects will be expounded in the next section and exemplified in Sec.~\ref{sec:5} for two common plasmonic metals: Al and Ag. As basically an atomic property, $\epsilon_p$ is not sensitive to temperature. The total current density now reads 
\begin{equation}
\mathbf{J}(\mathbf{x},t) = \mathbf{J}_b(\mathbf{x},t) + \mathbf{J}_s(\mathbf{x},t) + \mathbf{J}_p(\mathbf{x},t), \label{14}
\end{equation}
which determines the complete electrical responses of a semi-infinite metal. It is exactly $\mathbf{J}_s$ that is missing from the hydrodynamic/Drude model.  
 
\begin{figure*}
\begin{center}
\includegraphics[width=0.95\textwidth]{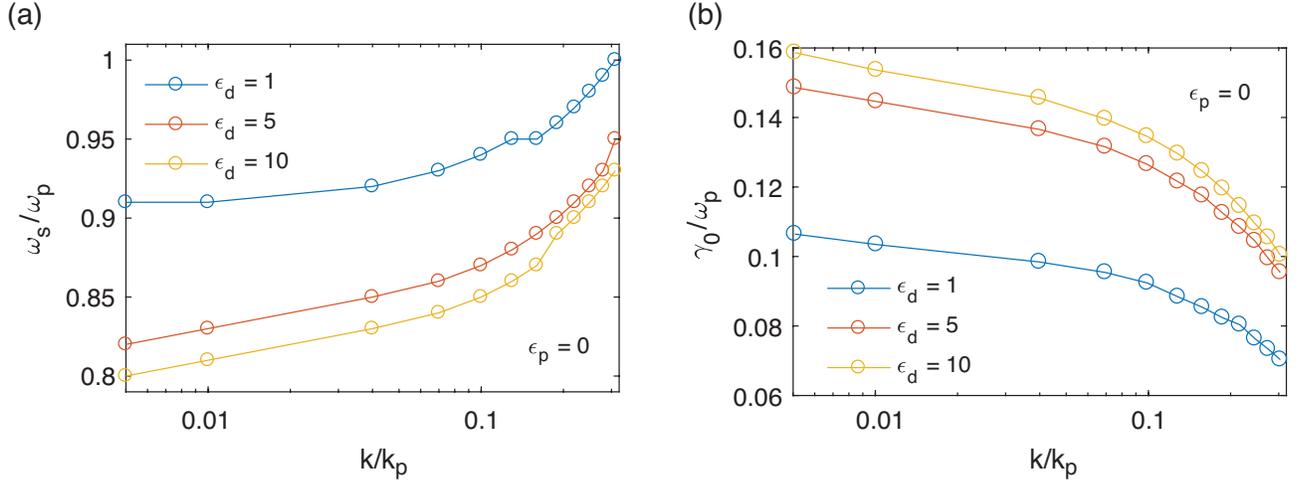}
\end{center}
\caption{Wavenumber dependence of $\bar{\omega} = \omega_s+i\gamma_0$, where $\omega_s$ and $\gamma_0$ denote respectively the frequency and the intrinsic rate of gain of the surface plasma waves (SPWs) supported on the interface between a metal and a dielectric with dielectric constant $\epsilon_d$. The results are obtained by numerically solving Eq.~(\ref{31}) in the absence of inter-band transition effects, i.e. for $\epsilon_p = 0$. Here $\omega_p$ is the characteristic frequency of the metal and $k_p = \omega_p/v_F$ with $v_F$ being the Fermi velocity. An upper cutoff $q_c=1.5k_p$ is used in the integration over $q$ in (\ref{31}). Solid lines are guides to the eye.\label{fig:f2}}
\end{figure*}  
 
\section{The fundamental equation for SPWs}
\label{sec:4}
The complete current densities (\ref{14}) can now be used to set up the dynamics of SPWs. We begin with the following equation of continuity for a semi-infinite system,
\begin{equation}
(\tau^{-1}+\partial_t)\rho(\mathbf{x},t) + \partial_{\mathbf{x}}\cdot\mathbf{J}(\mathbf{x},t) = -\delta(z)J_z(\mathbf{x}_0,t), \label{15}
\end{equation}
where $\mathbf{x}_0 = (x,y,z=0)$ and the r.h.s. arises due to the discontinuity in the current density across the surface. Using the quasi plane wave form, this equation is rewritten
\begin{equation}
-i\bar{\omega}\rho(z)+\nabla\cdot\mathbf{J}(z) = - J_z(0)\delta(z), \quad \nabla = (ik,\partial_y,\partial_z),\label{16}
\end{equation}
which can further be transformed as
\begin{equation}
(\mathcal{H}-\bar{\omega}^2)\rho(z) = S(z), \quad S(z) = i\bar{\omega}J_z(0)\delta(z).  \label{17}
\end{equation}
Here $\mathcal{H}$ is a linear operator on $\rho(z)$ defined by 
\begin{equation}
\mathcal{H}\rho(z) = -i\bar{\omega}\nabla\cdot\mathbf{J}(z), \label{18}
\end{equation}
which has to be complemented by the laws of electrostatics (\ref{19}). Equation (\ref{17}) admits of solutions giving either bulk plasma waves or SPWs, depending on whether $S(z)$ vanishes or not. SPWs correspond to solutions with non-vanishing source term, i.e. $S(z)\neq 0$ or equivalently $J_z(0)\neq 0$. Thus, SPWs originate from the current discontinuity caused by the surface. 

\begin{figure}
\begin{center}
\includegraphics[width=0.45\textwidth]{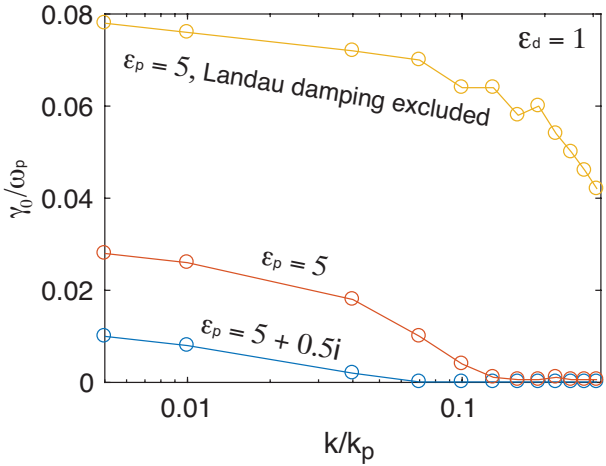}
\end{center}
\caption{Effects of inter-band absorption and Landau damping. The results are obtained by numerically solving Eq.~(\ref{31}) for $\epsilon_d = 1$. $q_c=1.5k_p$. Landau damping is excluded by taking only the real part of $\Omega(k,q;\bar{\omega})$ in Eq.~(\ref{31}). Solid lines are guides to the eye.\label{fig:f3}}
\end{figure}  

In line with the partition in Eq.~(\ref{14}), $\mathcal{H}$ is written as a sum of three portions: $\mathcal{H}_b$, $\mathcal{H}_s$ and $\mathcal{H}_p$, which are defined through Eq.~(\ref{18}) with the respective component of the current density in place of $\mathbf{J}$. Scrupulously following the same method as in our previous work~\cite{deng2017a}, we easily obtain
\begin{equation}
\mathbf{J}_b(z) = \frac{i}{\bar{\omega}}\frac{\omega^2_p}{4\pi}\mathbf{E}(z) +\mathbf{J}'(z), \label{a3}
\end{equation}
where $\omega_p = \sqrt{4\pi n_0e^2/m}$ is the characteristic frequency of the metal. The first term amounts to what is expected in the Drude model while
\begin{equation}
\mathbf{J}'(z) = \int^{\infty}_0 dq\frac{8\rho_q\mathbf{F}(\mathbf{k};\bar{\omega})}{k^2+q^2} \cos(qz) \label{a4}
\end{equation}
with $\mathbf{k} = (k,0,q)$ and $\mathbf{F}$ given by $$\mathbf{F} = \left(\frac{m}{2\pi\hbar}\right)^3\int d^3\mathbf{v}(-e^2f'_0)\mathbf{v}\sum^{\infty}_{l=2}\left(\frac{\mathbf{k}\cdot\mathbf{v}}{\bar{\omega}}\right)^l.$$ Note that $\mathbf{J}'(z)$ does not depend on $\epsilon_d$. Similarly, we obtain
\begin{eqnarray}
\mathbf{J}_s(z) &=& \left(\frac{m}{2\pi\hbar}\right)^3\int^{\infty}_0dq\frac{4k\rho_q}{k^2+q^2}\nonumber\\ &\quad&\quad \quad \int d^3\mathbf{v}\Theta(v_z)\left(-\frac{f'_0 e^2 \mathbf{v}}{v_z}\right) e^{i\frac{\tilde{\omega}z}{v_z}}\mathcal{L}(k,q;\bar{\omega}) \label{a5} 
\end{eqnarray}
with
\begin{equation}
\mathcal{L}(k,q,\bar{\omega})
 =\beta\frac{k(iv_x-v_z)}{kv_z+i\tilde{\omega}}-\frac{2(q^2v^2_z+\tilde{\omega}kv_x)}{\tilde{\omega}^2-q^2v^2_z}. \label{27}
\end{equation}
For $\epsilon_d = 1$, these expressions reduce to what we derived in our previous work~\cite{deng2017a}. 

Now we find
\begin{equation}
\mathcal{H}_b = \omega^2_p + \mathcal{H}', \quad \mathcal{H}'\rho(z) = -i\bar{\omega}\nabla\cdot\mathbf{J}'(z). \label{20}
\end{equation}
Note that $\mathcal{H}_b$ is not affected by the dielectric. $\mathcal{H}'$ is ignored in the Drude model but partially taken care of in the hydrodynamic model. This term gives rise to the dispersion of bulk plasma waves and Landau damping, see below. Using Eq.~(\ref{13}), $\mathcal{H}_p$ can be written as
\begin{equation}
\mathcal{H}_p = -4\pi i \bar{\omega} \sigma_p(\omega) = -\omega\bar{\omega}\epsilon_p(\omega). \label{21}
\end{equation}
As proved in our previous work, $\mathcal{H}_s$ is insignificant for small $kv_F/\omega_p$ and has negligible impact on both bulk waves and SPWs. It is hereafter left out. Thus, we arrive at
\begin{equation}
\mathcal{H} = \omega^2_p - \omega\bar{\omega}\epsilon_p(\omega) + \mathcal{H}'. \label{22}
\end{equation}
Notably $\mathcal{H}$ determines all the properties of bulk waves and displays no dielectric effects.~\cite{note2} For many materials, one may take $\epsilon_{p}$ as a constant and assume $\epsilon_{pi}\ll \epsilon_{pr}$. In such cases, Eq.~(\ref{22}) immediately yields in the Drude model for bulk waves the complex frequency $$\omega_b \approx \left(\omega_p/\sqrt{\epsilon_{\infty}}\right)\left(1-i\epsilon_{pi}/2\epsilon_{\infty}\right)$$ in the limit $\omega_p\tau\gg1$. Here $\epsilon_{\infty} = 1 + \epsilon_{pr}$. The imaginary part of $\omega_b$ signifies the damping due to inter-band absorption in bulk waves independent of $\epsilon_d$. 

Analogously, we can decompose the source term in three components, $S = S_b + S_s + S_p$ by the same token prescribed in Eq.~(\ref{14}). As shown in Ref.~\cite{deng2017a}, we have $J'_z(0)\equiv 0$. Thus, we obtain $J_{b,z}(0) + J_{p,z}(0) = (i/4\pi\bar{\omega})\left(\omega^2_p-\omega\bar{\omega}\epsilon_p(\omega)\right)E_z(0)$. In terms of $\rho_q$, we find from Eq.~(\ref{a2}) $$E_z(0) = -\beta\int^{\infty}_0 dq ~\frac{4k}{k^2+q^2}~\rho_q, \quad \beta = \frac{2\epsilon_d}{\epsilon_d+1}.$$ See that $E_z(0)$ is enhanced by the factor $\beta$ as a result of the induced charges in the dielectric. This is expected, since the surface sits exactly in between $\rho$ and $\rho_d$ and separates them spatially. Now we get
\begin{equation}
S_b(z)+S_p(z) = \delta(z) \int^{\infty}_0 dq~\frac{4~G_{bp}(k,\omega;\bar{\omega})}{k^2+q^2}~\rho_q,  \label{23}
\end{equation}
where the quantity 
\begin{equation}
G_{bp}(k,\omega,\bar{\omega}) = \beta\frac{k}{4\pi}\left(\omega^2_p-\omega\bar{\omega}\epsilon_p(\omega)\right) \label{24}
\end{equation}
is enhanced by the same factor $\beta$. Finally, the expression of $S_s$ can be obtained as
\begin{equation}
S_s(z) = \delta(z)  \int^{\infty}_0 dq~\frac{4~G_{s}(k,q;\bar{\omega})}{k^2+q^2}~\rho_q, \label{25}  
\end{equation}
where
\begin{equation}
G_s(k,q,\bar{\omega}) = i\bar{\omega}\left(\frac{m}{2\pi\hbar}\right)^3\int d^3\mathbf{v}\Theta(v_z)(-e^2f'_0)v_z\mathcal{L}(k,q,\bar{\omega}). \label{26}
\end{equation}
In the Drude model, where $\rho_q$ is taken as a constant with $\mathcal{H}'$ and the surface effects -- as contained in $S_s(z)$ -- left out, one obtains directly from Eq.~(\ref{17}) for SPWs the complex frequency $$\omega_\text{Drude} = \left(\omega_p/\sqrt{\epsilon_{\infty}+\epsilon_d}\right)\left(1-i\epsilon_{pi}/[2(\epsilon_{\infty}+\epsilon_d)]\right)$$ in the limit $\omega_p\tau\gg 1$. This expression, which is valid only if $\epsilon_p$ can be treated as a constant, suggests that although the presence of a dielectric does not affect bulk waves, it may well reduce the inter-band absorption in SPWs. 

Upon Fourier transforming Eq.~(\ref{17}) and utilizing (\ref{22}), we obtain 
\begin{equation}
\left(\Omega^2(k,q;\bar{\omega}) - \bar{\omega}^2\right)\rho_q = \bar{S}, \label{28}
\end{equation}
where $\Omega^2$ is the diagonal element of $\mathcal{H}$ in the $q$-representation and given by
\begin{equation}
\Omega^2 = \omega^2_p - \omega\bar{\omega}\epsilon_p(\omega) + \frac{4\pi\bar{\omega}\mathbf{k}\cdot\mathbf{F}}{\mathbf{k}\cdot\mathbf{k}} \label{29}
\end{equation}
See that $\Omega$ approaches a constant at large $|\mathbf{k}|$. In addition, 
\begin{equation}
\bar{S} = \int^{\infty}_0 dz ~S(z) \cos(qz) = \int^{\infty}_0 dq~\frac{4~G(k,q;\bar{\omega})}{k^2+q^2}~\rho_q \label{30}
\end{equation}
with $G = G_{bp} + G_s$. Obviously, the last term in Eq.~(\ref{29}) stems from $\mathcal{H}'$ in (\ref{22}) and is generally complex -- even in the collision\textit{less} limit when $\tau^{-1}$ is neglected -- due to a pole at $\bar{\omega} = \mathbf{k}\cdot\mathbf{v}$ in the integrand in $\mathbf{F}$. The imaginary part of $\Omega$ signifies the Landau damping. The real part of $\Omega$ approximates $\omega^2_p\left(1+\frac{3}{5}\frac{\mathbf{k}}{k_p}\cdot\frac{\mathbf{k}}{k_p}\right),$ where $k_p = \omega_p/v_F$, in the long wavelength limit. This is no more than the bulk wave dispersion relation already well known from the hydrodynamic theory. 

For SPWs the source is finite, i.e. $\bar{S}$ is finite. Combining Eqs.~(\ref{28}) and (\ref{30}), we arrive at~\cite{deng2017a,deng2017b}
\begin{equation}
1 = \int^{\infty}_0 dq ~\frac{4}{k^2+q^2} \frac{G(k,q;\bar{\omega})}{\Omega^2(k,q;\bar{\omega}) - \bar{\omega}^2}, \label{31}
\end{equation}
which constitutes the key equation for SPWs. To explicitly show how the intrinsic channel of gain can be generated by the surface effects, let us for the moment consider the collision\textit{less} limit and drop the dispersive term in Eq.~(\ref{29}), i.e. we put $\Omega^2 = \omega^2_p - \omega^2\epsilon_p$. Under this approximation, we may account for Landau damping by adding a negative imaginary part to $\omega_p$, namely $\omega_p \rightarrow \omega_p(1-i\eta_\text{Landau})$. In this way,  Eq.~(\ref{31}) becomes 
\begin{equation}
\frac{\omega^2_p}{1+\epsilon_p+\epsilon_d} - \omega^2 = \frac{1+\epsilon_d}{1+\epsilon_p+\epsilon_d}\int^{\infty}_0 dq~\frac{4~G_s(k,q;\omega)}{k^2+q^2}. \label{32}
\end{equation}
From Eqs.~(\ref{26}) and (\ref{27}), we find that~\cite{deng2017a}, to the linear order in $k$, $G_s \approx - \beta(k/2)(\omega^2_p/4\pi) -i(3\omega^2_p/16\pi)(qv_F/\omega)q$. With this we can rewrite Eq.~(\ref{32}) as
\begin{equation}
\omega^2_0 - \omega^2 = i~\frac{1+\epsilon_d}{1+\epsilon_p+\epsilon_d}\int^{\infty}_0 dq~\frac{4~\text{Im}\left[G_s(k,q;\omega)\right]}{k^2+q^2}, \label{33}
\end{equation}
where we have defined $$\omega_0 = \omega_p\sqrt{(1+\epsilon_d/2)/(1+\epsilon_p+\epsilon_d)},$$ which differs from $\omega_\text{Drude}$ because of the surface effects (see the expression of $\omega_\text{Drude}$ for comparison). 

For illustration, we again take $\epsilon_p$ as a constant. Substituting $\omega = \omega_0(1+i\eta_0)$ into Eq.~(\ref{33}) and assuming $\bar{\gamma}_0\ll 1$, we find 
\begin{equation}
\eta_0\approx - \frac{1}{2\omega^2_{s0}}\frac{1+\epsilon_d}{\epsilon_{\infty}+\epsilon_d}\int^{\infty}_0 dq~\frac{4~\text{Im}\left[G_s(k,q;\omega_{s0})\right]}{k^2+q^2}. \label{34}
\end{equation}
Here $\omega_{s0}=\text{Re}(\omega_0)$. Now that Im$\left[G_s(k,q;\omega_{s0})\right]<0$, we are led to $\eta_0>0$ concluding the existence of an intrinsic channel of gain. Expression (\ref{34}) implies that inter-band transitions have little effect on $\eta_0$, because the factor preceding the integral in this expression has no overall dependence on $\epsilon_p$. On the other hand, the presence of a dielectric enhances $\eta_0$. Taking into account Landau damping and inter-band absorption as well as electronic collisions, we obtain the rate of net gain as $\gamma = \gamma_0 - 1/\tau$, with 
\begin{equation}
\gamma_0 = \omega_{s0}\left(\eta_0 - \eta_\text{Landau} - \eta_\text{interband}\right), ~ \eta_\text{interband}=\frac{\epsilon_{pi}/2}{\epsilon_{\infty}+\epsilon_d}. \label{35}
\end{equation}
As the analysis presented in Sec.~\ref{sec:2} suggests, we expect to have $\eta_0$ surpassing $\eta_\text{Landau}$ and $\eta_\text{interband}$ put together so that $\gamma_0$ remains non-negative. Our numerical solutions confirm this, as demonstrated in what follows.   

We proceed to solve Eq.~(\ref{31}) numerically in the collision\textit{less} limit to find the SPW frequency $\omega_s$ and $\gamma_0$ and how they vary with $k$ and $\epsilon_d$. In the presence of inter-band effects, $\gamma_0$ generally depends on the electronic collision rate, but such dependence is insignificant and therefore ignored here. In Eq.~(\ref{31}), the integral over $q$ extends to infinity. In practice, however, there is a natural cutoff $q<q_c$, namely $\rho(z)$ cannot vary significantly over a distance of the order of a lattice constant $a$ of the metal; otherwise, the jellium model (and hence our theory) would break down. Thus, $q_c\sim a^{-1}$. For metals, this means $q_c \sim k_F \sim k_p = \omega_p/v_F$. In all the numerical results to be presented, we have chosen $q_c = 1.5k_p$, which is sufficiently large for the metals considered in this work. We also point out that, as the present theory has disregarded retardation effects, the calculations should be taken with a grain of salt for very small $k$, i.e. $k\ll \omega_s/c \approx k_pv_F/c \sim 0.01k_p$, where $c$ is the speed of light in vacuum. 

The numerical results are displayed in Figs.~\ref{fig:f2} and \ref{fig:f3}. In Fig.~\ref{fig:f2}, we show $\bar{\omega} = \omega_s+i\gamma_0$ as a function of $k$ for various $\epsilon_d$ but without inter-band transition effects ($\epsilon_p=0$). As seen in Fig.~\ref{fig:f2} (a), in agreement with what is suggested of the expression of $\omega_{s0}$, increasing $\epsilon_d$ leads to smaller $\omega_s$. Note that $\omega_s$ is considerably larger than what would be obtained from $\omega_\text{Drude}$. On the other hand, $\gamma_0$ increases with increasing $\epsilon_d$, as seen in Fig.~\ref{fig:f2} (b), in accord with Eq.~(\ref{34}). The effects of inter-band absorption and Landau damping are illustrated in Fig.~\ref{fig:f3}. Here we plot $\gamma_0$ for $\epsilon_d=1$ under several circumstances as described in the figure. We see that inter-band transitions can strongly diminish $\gamma_0$ in two ways, as can be deduced from Eqs.~(\ref{34}) and (\ref{35}). First, there is the screening effect (the curve labelled $\epsilon_p=5$). This leads to smaller $\omega_{s0}$ and hence smaller $\gamma_0$, while leaving $\eta_0$ unaffected. Second, inter-band absorption further reduces $\gamma_0$ (see the curve with $\epsilon_p=5+0.5i$). As for Landau damping, what is interesting is the observation that it is not only sizable but also almost $k$ independent, in contrast to that with bulk plasma waves. This feature has already been pointed out in Ref.~\cite{khurgin} and can be easily understood: the highly localized nature of SPWs makes virtually all $\rho_q$ component present and contribute to the imaginary part of $\Omega$, regardless of the value of $k$ [see Eq.~(\ref{29})]. Finally, we observe that $\gamma_0$ increases as $k$ decreases. In regard to this feature, we should mention a size effect: in the case of metal films of thickness $d$, this increase will be terminated for $k<1/d$, where $\gamma_0$ would approach zero quickly~\cite{deng2017b} as $k$ further decreases thence.

\begin{figure*}
\begin{center}
\includegraphics[width=0.97\textwidth]{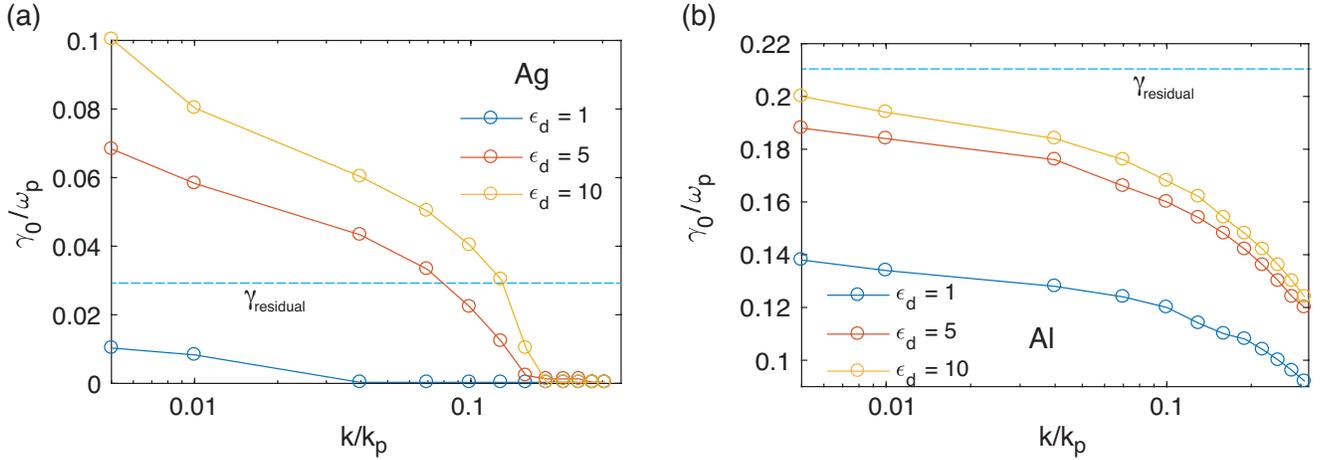}
\end{center}
\caption{The intrinsic channel of gain for SPWs in Ag and Al at various $\epsilon_d$. In both metals, the gain rate $\gamma_0$ is enhanced upon replacing the vacuum with a dielectric. At long wavelengths $\gamma_0$ can even be made in excess of $\gamma_\text{residual}$ -- the value of $\tau^{-1}$ at zero temperature -- in Ag, whereby allowing the plasmonic losses to be completely compensated, see Fig.~\ref{fig:f5}. The caveats of these calculations are discussed in the main text. $q_c=1.5k_p$. Solid lines are guides to the eye. \label{fig:f4}}
\end{figure*} 

\section{The cases with $\mbox{Ag}$ and $\mbox{Al}$}
\label{sec:5}
We have so far revealed the existence of an intrinsic channel of gain for SPWs. Now we discuss the possibility of using it to fully compensate for the losses in real metals. As shown above, the ultimate losses to be circumvented is that due to electronic collisions. Here we argue that the rate of net losses, i.e. $(\tau^{-1} - \gamma_0)$ can be made to vanish in some metals by simply cooling down the system, based on the observation that $\gamma_0$ is insensitive to temperature variations~\cite{note3} while $\tau^{-1}$ can be reduced upon cooling. If the residual collision rate $\gamma_\text{residual}$, that is, $\tau^{-1}$ at zero temperature, is less than $\gamma_0$, then at certain temperature $T^*$ the losses can be fully compensated for. In the presence of SPWs $\gamma_\text{residual}$ is finite even in defect-free metals. In ultra-pure samples, it may be estimated by the following formula~\cite{beach1977}:
\begin{equation}
\gamma_\text{residual} \approx \frac{2\nu_0}{5} + \frac{\omega_p}{4\pi^2}\left(\frac{\omega_s}{\omega_p}\right)^2, \label{36}
\end{equation}
where the first contribution arises from scattering with phonons and the second from electron-electron scattering. It is noted that the second term generally underestimates the electron-electron rate for noble metals and Al by a few times~\cite{beach1977}. The coefficient $\nu_0\sim k_BT_D/\hbar$ can be determined from the slope of the phononic part of the D.C. conductivity of metals at temperatures higher than the Debye temperature $T_D$. Equation (\ref{36}) shows that $\gamma_\text{residual}$ may well amount to quite a few percentages of $\omega_p$ for metals. At finite temperatures, $\tau^{-1}$ may be evaluated by
\begin{equation}
\tau^{-1} \approx \gamma_\text{residual} + 4\nu_0\left(\frac{T}{T_D}\right)^5\int^{T_D/T}_0\frac{y^4dy}{e^y-1} + \omega_p\left(\frac{k_BT}{\hbar\omega_p}\right)^2. \label{37}
\end{equation}
The last term here is obviously negligible in the temperature region of interest but the second could reach a sizable fraction of $\omega_p$ at room temperature. As demonstrated in the last section, under optimal conditions where the inter-band absorption is not pronounced, $\gamma_0$ can also be as large as a tenth of $\omega_p$ and is therefore capable of significantly counteracting $\tau^{-1}$. In what follows, we assess this expectation in Ag and Al, which are amongst the most experimented plasmonic materials. 

In silver electronic transitions involving the \textit{s} and \textit{d} bands have a dramatic effect on the properties of SPWs~\cite{marini2002,liebsch1993}, leading to $\hbar\omega_{s}\approx 3.69~$eV and $\hbar\omega_b \approx 3.92~$eV at long wavelengths, both lying far below the characteristic frequency $\hbar\omega_p = 9.48~$eV. Experimentally~\cite{beach1977,note4}, it was found that $\hbar\nu_0\sim 0.1~\text{eV}\sim 0.01\hbar\omega_p$, in consistency with the value of $T_D \approx 220~$K for Ag. The electron-electron scattering rate according to Eq.~(\ref{36}) would be less than one percent of $\omega_p$, while experimental measurements and more accurate expressions~\cite{beach1977} place it about $0.02\omega_p$. As such, we may reasonably take $\gamma_\text{residual}\sim0.03\omega_p$. The SPW damping rate (i.e. $-\gamma$) can be directly read out from -- for example -- the line shape of the electron energy loss (EEL) spectra. The temperature dependence of $-\gamma$ has been recorded on a high-quality single Ag crystal by means of EEL spectroscopy~\cite{rocca1992}. The data indicates that at zero temperature $-\gamma$ counts for less than one percent of $\omega_p$, a value a few times smaller than the as-projected $\gamma_\text{residual}$. This paradox is resolved thanks to the existence of the intrinsic channel of gain. The estimate suggests that $\gamma_0$ has substantially compensated for the losses, i.e. $\gamma_0 \sim \gamma_\text{residual}$, a statement well borne out in our theory, see what follows. 

To compute $\gamma_0$ by the theory, $\epsilon_p(\omega)$ must be supplied in Eq.~(\ref{31}). So far as we are concerned, no direct measurement of this quantity has been performed for Ag. Combining a semi-quantum (Lorentz oscillator) model and ellipsometry as well as transmittance-reflectance measurements, Raki\'{c} \textit{et al.}~\cite{rakic1998} employed K-K analysis and prescribed a parametrized dielectric function -- $\hat{\epsilon}^{(b)}_r(\omega)$ in their notation -- for the inter-band contribution. In this work, we use their fitting as an input for $\epsilon_p(\omega)$ but with a number of caveats. Firstly, their dielectric function was deduced from measurements assuming the conventional electromagnetic responses without any surface effects, i.e. the effects contained in $\mathbf{J}_s$. These effects, however, should be considered when analyzing ellipsometry and reflectance spectra. A future study will be made to address this issue systematically. Secondly, their function very poorly reproduces the EEL function and the reflectance spectra, especially near the SPW frequency of interest. Thirdly, their function does not give an accurate partition into inter-band and intra-band contributions, e.g. $\omega_p\approx 8$eV was used rather than the widely agreed $9.48$eV~\cite{marini2002}, which may overestimate the inter-band transition effects. Finally, their function is defined only for real frequency, and thus in general not suitable for $\epsilon_p(\omega)$, where $\omega$ is complex. To remedy this, we substitute $\hat{\epsilon}^{(b)}_r[\text{Re}(\omega)]$ for $\epsilon_p(\omega)$, which should be reasonable if $\gamma/\omega_s\ll 1$. 

\begin{figure}
\begin{center}
\includegraphics[width=0.47\textwidth]{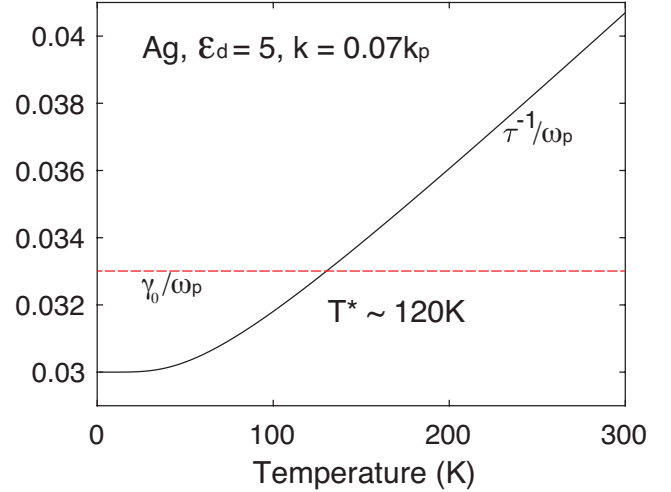}
\end{center}
\caption{The possibility of compensating for the plasmonic losses in Ag with the intrinsic channel of gain. In this metal, $\gamma_\text{residual}$ could be made smaller than $\gamma_0$ at long wavelengths by a dielectric. A critical temperature $T^*$ exists where $\gamma_0 = \tau^{-1}$. Cooling down the system toward $T^*$ reduces the losses to a vanishingly small level. Here $\tau^{-1}$ is calculated by Eq.~(\ref{37}) with $T_D=220$K and $\hbar\nu_0 = 0.1$eV. \label{fig:f5}}
\end{figure} 

The results are displayed in Fig.~\ref{fig:f4} (a), where $\gamma_0$ is exhibited as a function of $k$ at various values of $\epsilon_d$. While $\gamma_0$ for SPWs supported o a pristine Ag surface (i.e. $\epsilon_d = 1$) is negligibly small and way below $\gamma_\text{residual}$, by using a dielectric $\gamma_0$ can be significantly enhanced at long wavelengths beyond $\gamma_\text{residual}$. This trend is consistent with Eq.~(\ref{34}). The fact that $\gamma_0$ can be made higher than $\gamma_\text{residual}$ suggests the possibility of compensating for the plasmonic losses completely in Ag. The situation is shown in Fig.~\ref{fig:f5}. By cooling down the metal toward a critical temperature $T^*$, one can diminish the net losses as much as desired. 

Inter-band transitions in Al are widely considered less pronounced than in Ag. Nevertheless, their presence can still be felt, e.g. in the difference between the values of $\hbar\omega_p \approx 12.6$eV and $\hbar\omega_b \approx 15.3$eV. These numbers were obtained by density functional theory calculations~\cite{lee1994} and experimental fitting~\cite{rakic1995}. In addition, $\hbar\omega_s\approx 10.7$eV~\cite{powell1959,knight2014}. $\gamma_\text{residual}$ may be deduced from the experimental measurements performed by Sinvani \textit{et al.}~\cite{sinvani1981} and others~\cite{ribot1979}. These authors measured the low temperature dependence of the d.c. resistivity $\rho$ of Al. Their data shows that $\rho \approx \rho_0 + AT^2 +B(T/T_D)^5$, where $\rho_0$ stems from impurity and lattice dislocation scattering while $A$ and $B$ are constants characterizing electron-electron scattering and electron-phonon scattering, respectively. Analyzing the data, the authors found that $A\approx 0.21$p$\Omega\cdot$cm$/$K$^2$ (a lower bound) and $B\approx 4.9\cdot 10^4\mu\Omega\cdot$cm for $T_D=430$K. From this we obtain $\nu_0\approx0.18\omega_p$ and the residual electron-electron scattering rate -- a few times larger than what would be obtained with Eq.~(\ref{36})~\cite{beach1977} --  approximating $0.14\omega_p$, yielding $\gamma_\text{residual}\approx 0.21\omega_p$. It is noted that this value is comparable to the width ($\sim 1.5$eV, nearly $0.12\omega_p$) of the EEL peak for Al corresponding to the bulk plasma waves~\cite{powell1959}, for which no intrinsic gain channel exists. The as-obtained $\nu_0$ (and hence $\gamma_\text{residual}$) represents probably an overestimate~\cite{note5}. 

To compute $\gamma_0$, we again resort to the fitting function constructed by Raki\'{c} \textit{et al.}~\cite{rakic1995} and have it in place of $\epsilon_p(\omega)$ in our theory, in the same way as we did in the case of Ag. It goes without saying that the same caveats should be kept in mind. The results are shown in Fig.~\ref{fig:f4} (b). As expected, $\gamma_0$ is comparable to that in the absence of inter-band transitions [see Fig.~\ref{fig:f2} (b)], as these transitions are weak in Al. As is with Ag, the intrinsic channel of gain for SPWs in Al can also be fortified -- but to a lesser extent -- by a dielectric. Nevertheless, the enhanced $\gamma_0$ still falls short of $\gamma_\text{residual}$ unless for very long wavelengths where retardation effects need to be properly accounted for. 

\section{Discussions and Conclusions}
\label{sec:6}
We have presented a SPW theory including both inter-band and dielectric effects. As asserted in our previous work~\cite{deng2017a,deng2017b,deng2016}, we have shown that SPWs must possess an intrinsic channel of gain due to genuine surface effects. We found this channel to be affected by the inter-band and dielectric effects: it can be significantly boosted in the presence of a dielectric whereas much weaker with inter-band transitions. To demonstrate that it can be harnessed to beat the energy losses, we have applied the theory to two real materials: Ag and Al. We found that, with a dielectric the gain channel can be sufficiently strengthened so that all the plasmonic losses could be compensated by tuning the system toward a critical temperature from above. This prediction is ready for experimental study in many ways, e.g. using EEL spectroscopy or measuring the SPW propagation distance. In case a thin metal film is used, one should bear in mind the size effect~\cite{deng2017b}:  the film thickness should much exceed the SPW wavelength or the intrinsic channel of gain would be diminished to a negligible level. 

Our analysis also reveals that, in general the residual loss rate $\gamma_\text{residual}$ amounts to quite a few percentages of the characteristic frequency $\omega_p$ of the metal. For example, in Ag it is around $3\%$ while in Al about $21\%$. However, the observed loss rate~\cite{tsuei1991} even at room temperature in those metals stays far below $\gamma_\text{residual}$. We take this as an evidence for the existence of the intrinsic channel of gain $\gamma_0$. Indeed, our calculations have shown that $\gamma_0$ well compares to $\gamma_\text{residual}$ in magnitude and is therefore expected to have substantially compensated for the latter, whereby returning the observed low loss rate. In conjunction with this point, we may note that the loss rate of bulk plasma waves differs from that of SPWs primarily because of the lack of the intrinsic channel in the former. As such, bulk waves should be generally much more lossy than SPWs, an observation that seems in consistency with experience. 

Our theory is built within the regime of linear electrical responses of metals residing in the state of the Fermi sea. The fact that $\gamma_0$ could be made bigger than $\tau^{-1}$ under certain circumstances -- e.g. for temperatures below $T^*$ -- means an instability of the Fermi sea. Upon entering such circumstances, the metals are expected to undergo a phase transition into a new state other than the Fermi sea. This state should be stable and the electrical responses here may not be captured by our calculations. We consider it an urgent issue to clarify the nature of this transition in the future.  

The results reported in this work should be of broad interest to the researchers working in plasmonics, surface science and condensed matter physics. We hope that the experimentalists will find the results fascinating enough to put their hands on them. 

\textbf{Acknowledgement} -- This work is not supported by any funding bodies.


\begin{thebibliography}{9}
\bibitem{note2}
Inclusion of $\mathcal{H}_s$ would lead to negligible dependence on $\epsilon_d$.
\bibitem{note3}
The correction should be of the order of $(k_BT/\varepsilon_F)^2$ and hence negligible in the region of temperatures of interest.
\end{thebibliography}
\end{document}